\documentclass[usenatbib]{mn2e}

\usepackage{natbib}
\usepackage{amsmath}
\usepackage{amssymb}
\usepackage{graphicx}
\usepackage{url}
\newcommand{\spider}{{\sc Spider}}
\newcommand{\ebex}{{\sc EBEX}}

\newcommand{\biceptwo}{{\sc Bicep2}}
\newcommand{\WMAP}{{\sc WMAP}}
\newcommand{\FGPol}{{\sc FGPol}}

\newcommand{\Archeops}{{\sc Archeops}}

\newcommand{\healpix}{{\sc healpix}}

\newcommand{\sech}{\mathrm{\,sech\,}}
\newcommand{\nn}{\nonumber}
\newcommand{\ud}{{\mathrm{d}}}
\newcommand{\vect}[1]{\ensuremath{\boldsymbol{#1}}}

\newcommand{\abs}{{\sc ABS}}

\newcommand{\polarbear}{{\sc POLARBEAR}}

\newcommand{\keck}{{\sc Keck}}

\newcommand{\beginfigure}{\begin{figure*}}
\newcommand{\efigure}{\end{figure*}}
\graphicspath{{./pdf/}{./eps/}}

\title[Modelling Polarisation Foregrounds]{Modelling the Polarisation of Microwave Foreground Emission on Large
  Angular Scales}
\author[C.~N.~Clark, C.~R.~Contaldi, and C.~J.~MacTavish]{
  C.~N.~Clark$^{1}$\thanks{E-mail:
    caroline.clark05@imperial.ac.uk},  C.~R.~Contaldi$^{1}$, and C.~J.~MacTavish$^{2}$\\
  $^{1}$Theoretical Physics, Blackett Laboratory, Imperial
  College, London, UK\\
  $^{2}$Kavli Institute for Cosmology, University of Cambridge, Cambridge, UK}
\begin{document}

\pagerange{\pageref{firstpage}--\pageref{lastpage}} \pubyear{2012}

\maketitle

\label{firstpage}
\begin{abstract}
  Templates for polarised emission from Galactic foregrounds at
  frequencies relevant to Cosmic Microwave Background (CMB)
  polarisation experiments are obtained by modelling the Galactic
  Magnetic Field (GMF) on large scales.  This work extends the results
  of~\citet{2012MNRAS.419.1795O} (hereafter \FGPol I) by including
  polarised synchrotron radiation as a source of foreground
  emission. The polarisation direction and fraction in this
  calculation are based solely on the underlying choice of GMF model
  and therefore provide an independent prediction for the polarisation
  signal on large scales. Templates of polarised foregrounds may be of
  use when forecasting effective experimental sensitivity.  In turn,
  as measurements of the CMB polarisation over large fractions of the
  sky become routine, this model will allow for the data to constrain
  parameters in the, as yet, not well understood form of the
  GMF. Template foreground maps at a range of frequencies can be
  downloaded from the on-line
  repository~\footnote{\url{http://www.imperial.ac.uk/people/c.contaldi/fgpol}}.
 \end{abstract}


\section{Introduction}
Currently operating or upcoming CMB experiments such as \ebex\
\citep{2010SPIE.7741E..37R}, \spider\ \citep{2010SPIE.7741E..46F},
\polarbear\ \citep{2010arXiv1011.0763T}, \keck\
\citep{2011arXiv1104.5516S} and \abs\ \citep{2010arXiv1008.3915E} will routinely
reach the sensitivity in polarisation required to detect the curl-type
pattern ($B$-mode) predicted by the simplest models of inflation
\citep{Dodelson:2009kq}. The predicted amplitude of this signal
however is comparable or below the predicted signal of foreground
polarisation over all observationally relevant frequencies and over
most of the sky \citep{2011ApJS..192...15G}. For polarisation in particular the
foreground signal is dominated by synchrotron emission at low
frequencies ($\lesssim$ 100 GHz) and thermal dust emission at high
frequencies ($\gtrsim$ 100 GHz).

Given the high levels of polarised foregrounds the mission planning
for these and future experiments requires a detailed study of sky coverage
to optimise sensitivity to the $B$-mode signal.  The
impact of the trade off between larger sky coverage, depth of
observation and the `cleanliness' of or lack of Galactic foregrounds
in a patch of sky must all be considered.  Regardless of how clean the
final observed sky patch, some level of foreground removal will be
required for all experiments.  Realistic foreground templates based on
models and/or observations of the polarisation direction and
amplitude of foregrounds are very useful when carrying out this
work. However, reliable templates of polarised foregrounds at the
frequencies relevant to CMB observations have been hard to come by and
only recently, with  Wilkinson Microwave Anisotropy
Probe (\WMAP) $K$-band observations have reliable estimates
of synchrotron polarisation on large angular scales been made.

Little other polarization data exists at the frequencies of interest
for future CMB experiments (with frequency bands ranging from $90$ GHz
up to $450$ GHz), hence it is necessary to model foreground emission
by extrapolating the information from existing data.  This paper
builds on previous work presented in \FGPol I, which described a model
for foreground emission due to interstellar dust in the Galaxy.  The
dust model was introduced by \cite{Danielthesis:2009} and first
applied in~\citet{2011ApJ...738...63O} for the purpose of studying the
impact of polarised foreground dust on \spider's ability to detect
$B$-mode polarisation.  \FGPol I gives a detailed explanation of the
dust model and presents a number of full-sky template maps at various
frequencies.  A complete model of polarised foreground emission must
also include the effect of synchrotron emission. This is particularly
important for low frequency observations i.e. below the CMB
`sweet-spot' at 100 GHz. This will be the focus of the work reported
here. Additional components due to spinning dust and free-free
emission are not thought to give a significant signal in polarisation
and are omitted in our modelling (see for example
\citet{2011arXiv1108.0205M} and \citet{2011ApJ...729...25L}).

Synchrotron emission, generated by the gyration of cosmic ray
electrons in the Galactic magnetic field (GMF), is intrinsically
polarised and constitutes the main polarised foreground at lower
frequencies \citep{2007ApJS..170..335P}.  While the emission from
thermal dust is expected to be higher than synchrotron emission above
around $90$\,GHz, the signal from synchrotron is also not negligible.
With the addition of synchrotron, this paper provides a complete model
of polarised microwave foreground emission on large angular scales.
Here a detailed explanation of the synchrotron model is given and
full-sky template maps are presented. As with previous work the model
includes a three dimensional description of the Galactic magnetic
field on both large and small spatial scales.  Both polarisation
amplitude and angle are modelled internally and the templates are
scaled such that the polarisation amplitude corresponds to a nominal
value when averaged over the maps.

This paper is organized as follows.  Section~\ref{sec:synch} briefly
reviews the mechanism for synchrotron
emission. Section~\ref{sec:synchmodel} describes the synchrotron model
including the underlying GMF, the choice of total intensity template,
and the cosmic ray density distribution and line-of-sight integration
method required to evaluate the final Stokes parameters maps.  In
Section~\ref{sec:maps} examples of template maps produced using this
model are described and in Section~\ref{sec:wmap_comp} the templates
are are compared with the \WMAP\ estimated foreground maps.
Section~\ref{sec:suborb} compares the level of foreground polarisation
in the templates in coverage areas of a sample of currently planned or
operating sub-orbital experiments. We conclude with some discussion in
Section~\ref{sec:conclude}.

\begin{figure*}
 \begin{center}
   \begin{tabular}{c}
     \makebox[3in][c]{\includegraphics[width=8cm,angle=180,trim=0cm 1cm 0cm 1cm,clip]{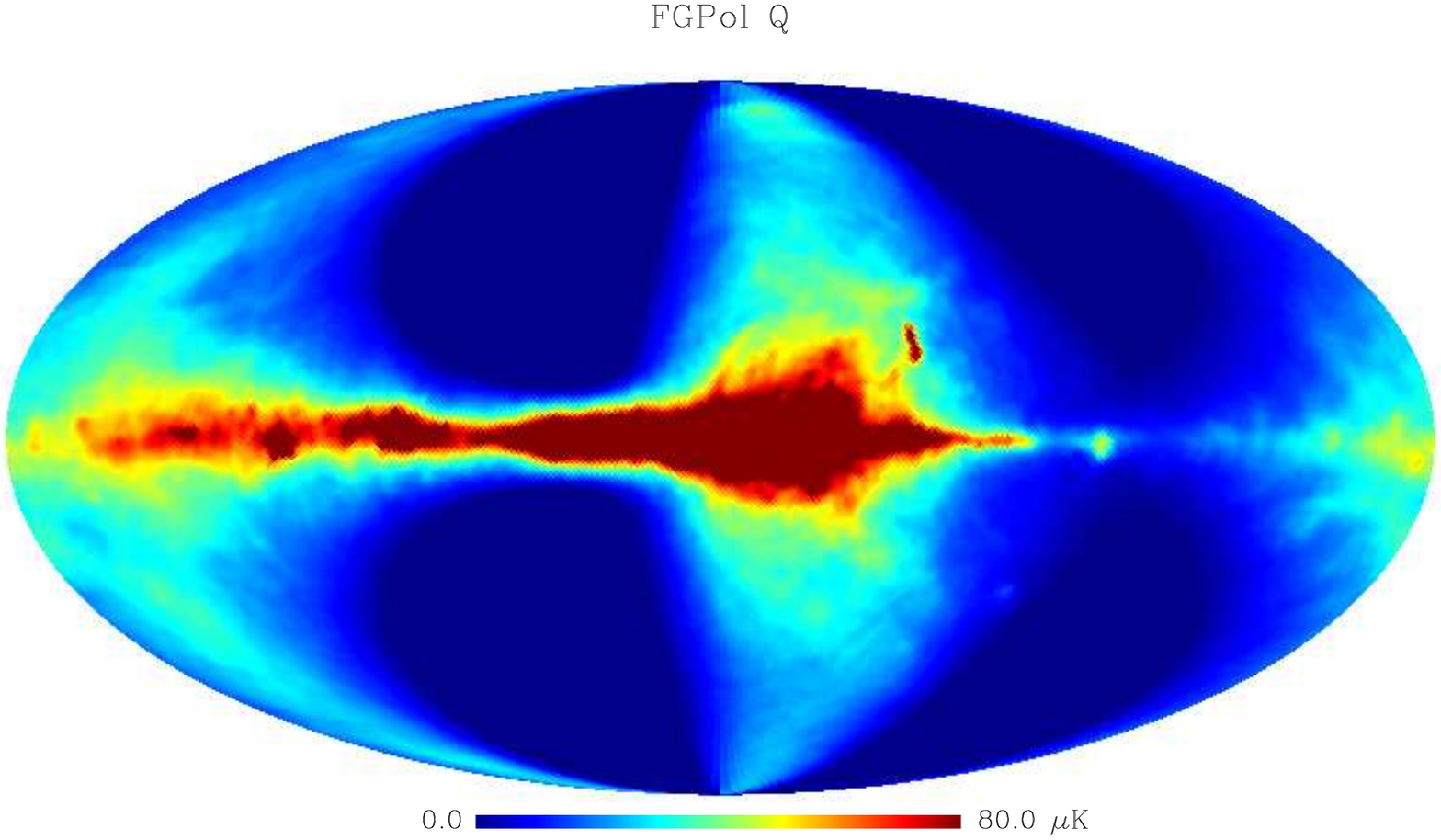} 
\includegraphics[width=8cm,angle=180,trim=0cm 1cm 0cm 1cm,clip]{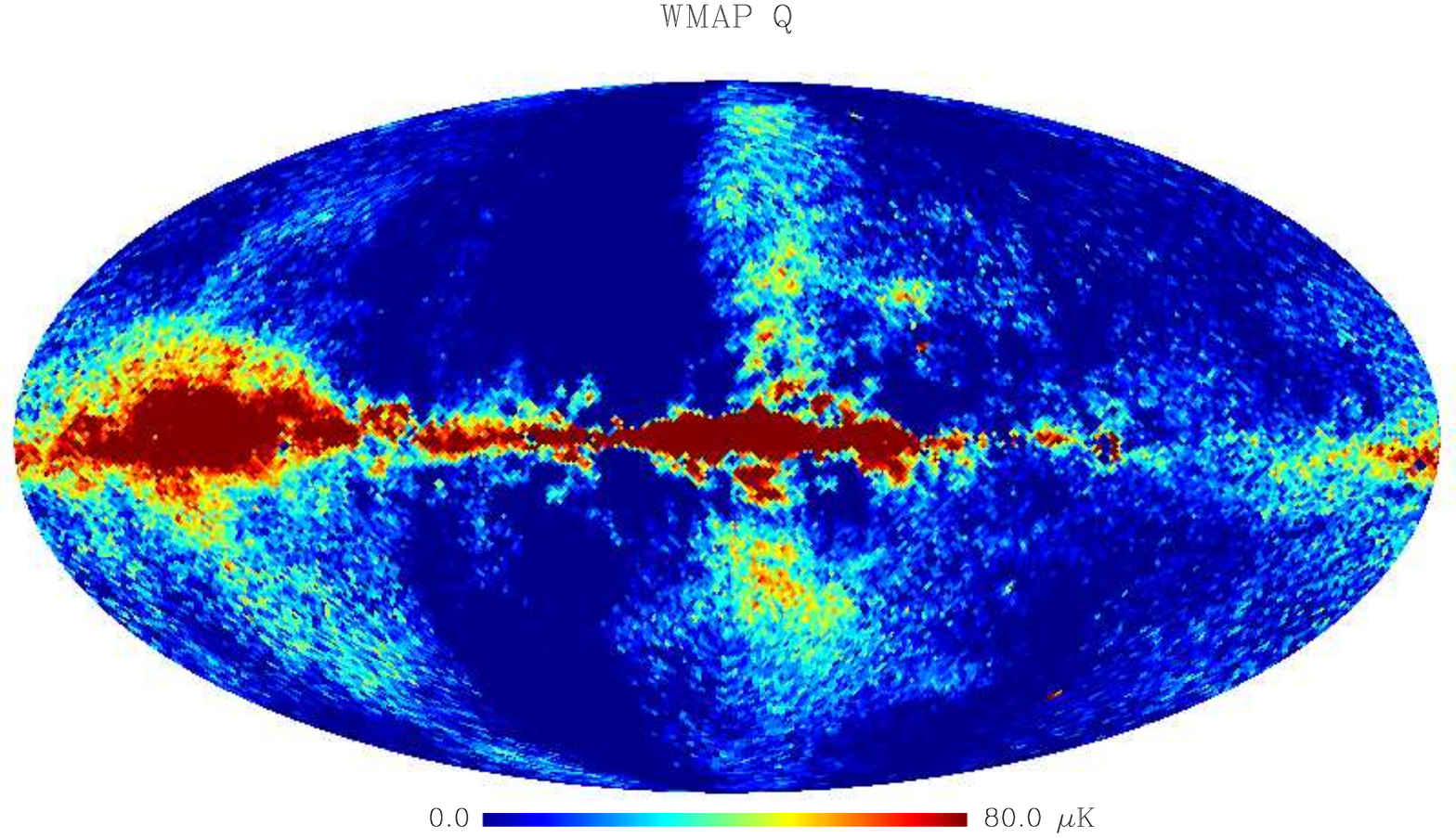}} \\
 \makebox[3in][c]{\includegraphics[width=8cm,angle=180,trim=0cm 1cm 0cm 1cm,clip]{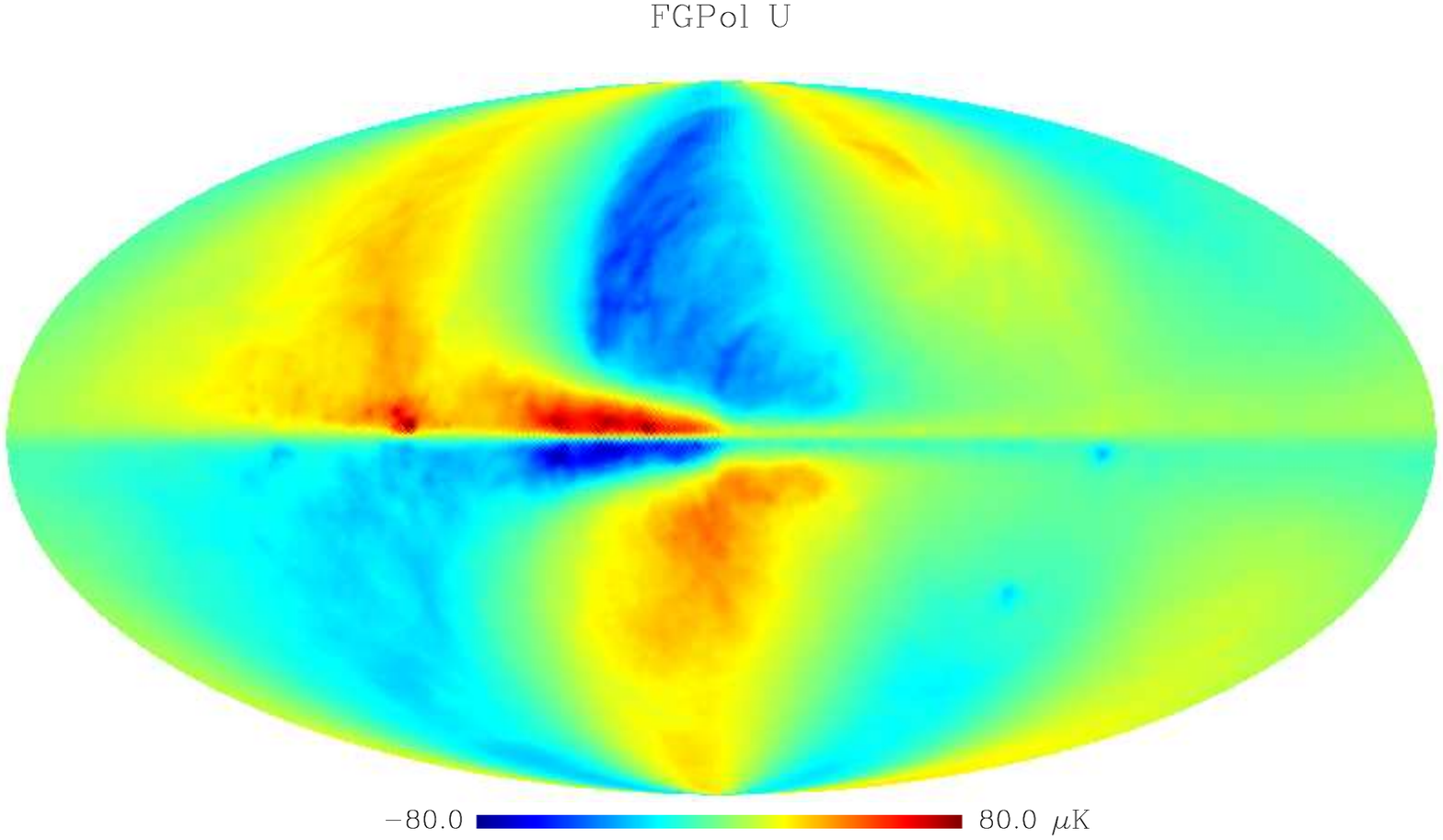} 
\includegraphics[width=8cm,angle=180,trim=0cm 1cm 0cm 1cm,clip]{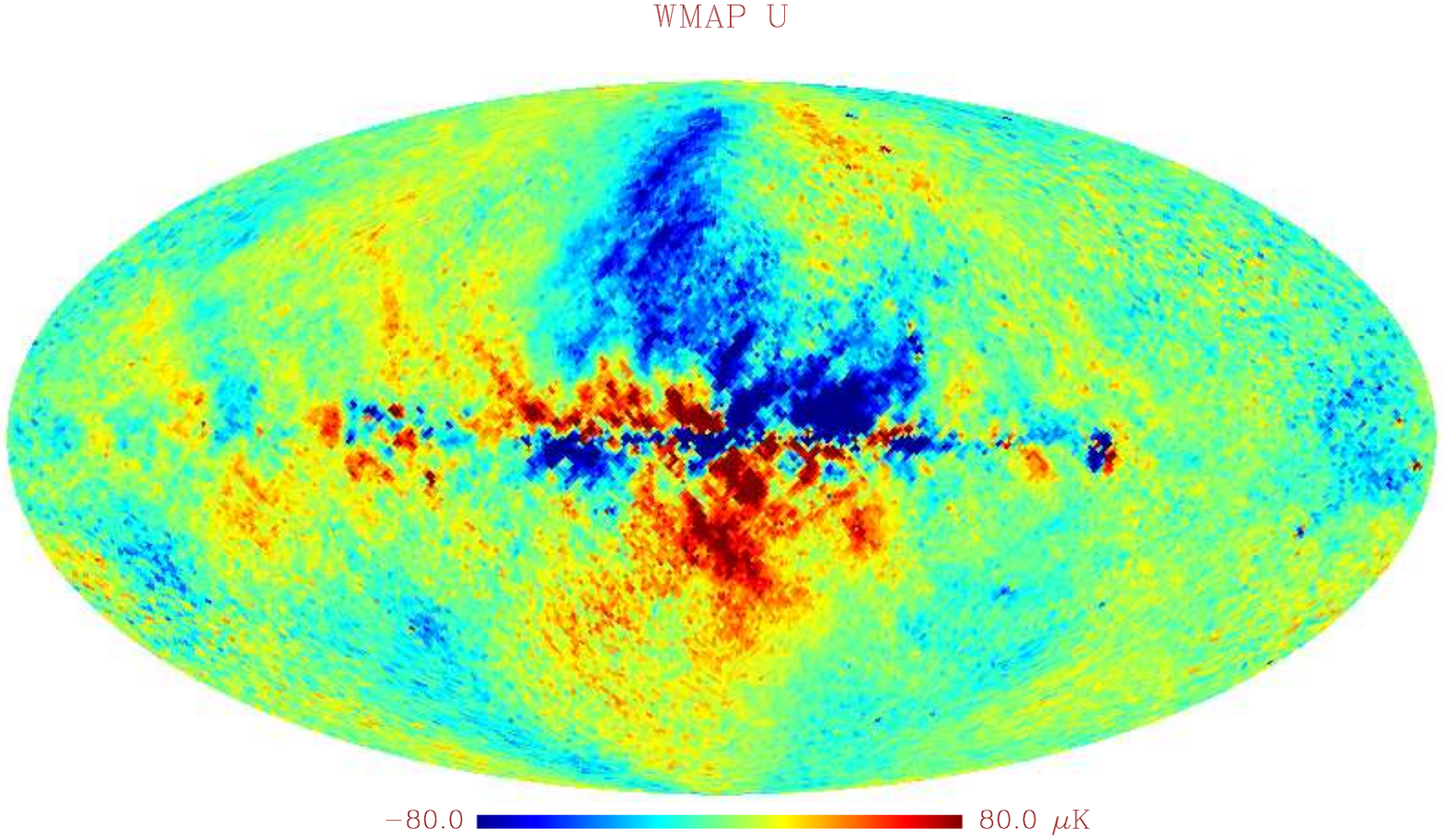}} \\
 \end{tabular}
 \caption{From top to bottom, $Q$ and $U$ Stokes parameter template
   maps, displayed in Galactic co-ordinates for {\sl left}: \FGPol\
   model of synchrotron emission at $23$\,GHz for the LSA GMF
   model. These were generated at $N_{\rm side}=1024$ map resolution
   based on $N_{\rm side}^{\rm P}=128$ line-of-sight resolution. The
   maps are smoothed to $1^\circ$ and downgraded to $N_{\rm
     side}=64$. {\sl right}: \WMAP\ MCMC synchrotron map for
   comparison. Units are $\mu K$ antenna temperature.}
 \label{fig:iqu}
 \end{center}
 \end{figure*}
   
\begin{figure*}
 \begin{center}
 \begin{tabular}{c}
     \makebox[3in][c]{\includegraphics[width=8cm,angle=180,trim=0cm 1cm 0cm 1cm,clip]{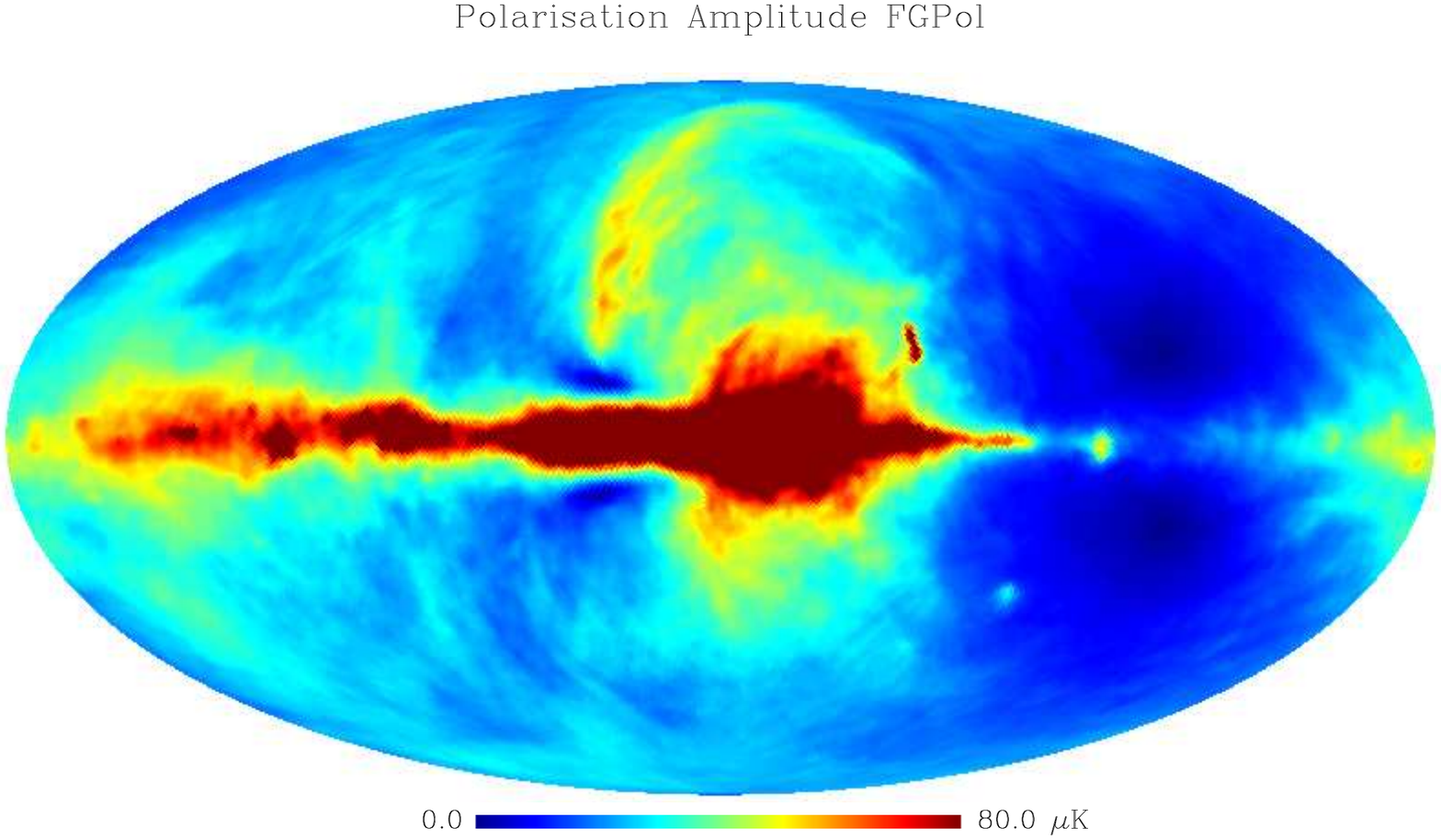} \includegraphics[width=8cm,angle=180,trim=0cm 1cm 0cm 1cm,clip]{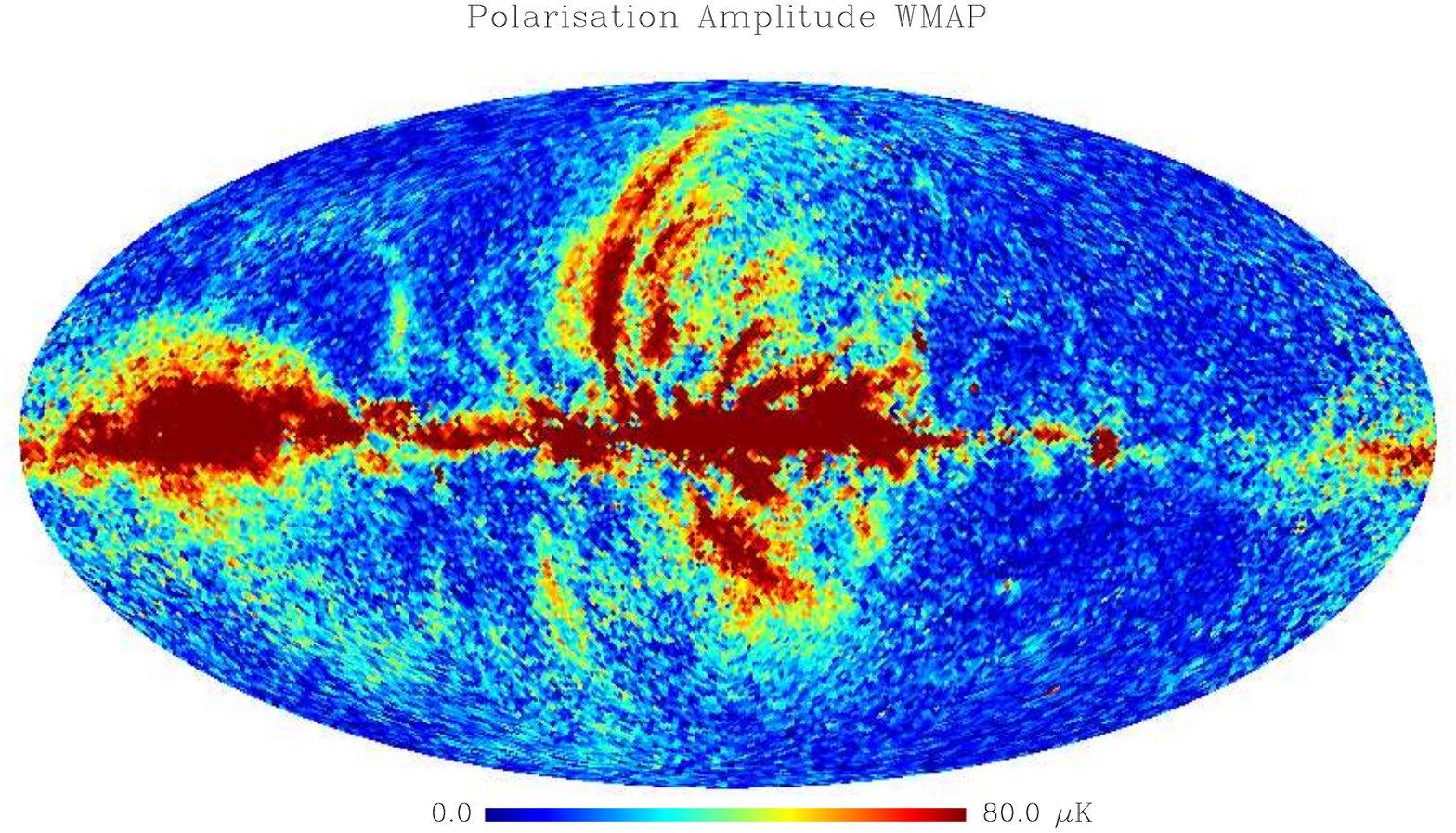}} \\
     \makebox[3in][c]{\includegraphics[width=8cm,angle=180,trim=0cm 1cm 0cm 1cm,clip]{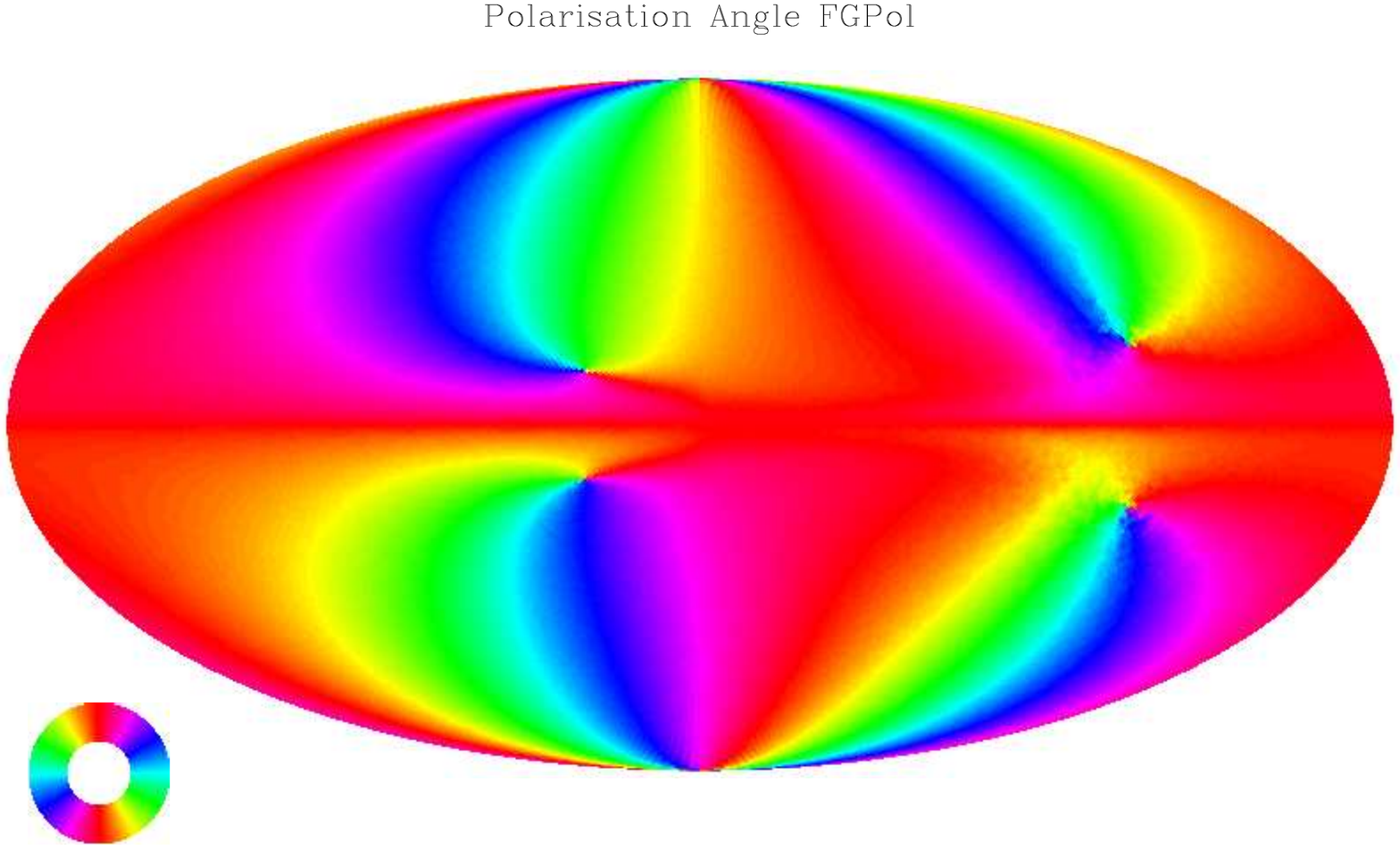} \includegraphics[width=8cm,angle=180,trim=0cm 1cm 0cm 1cm,clip]{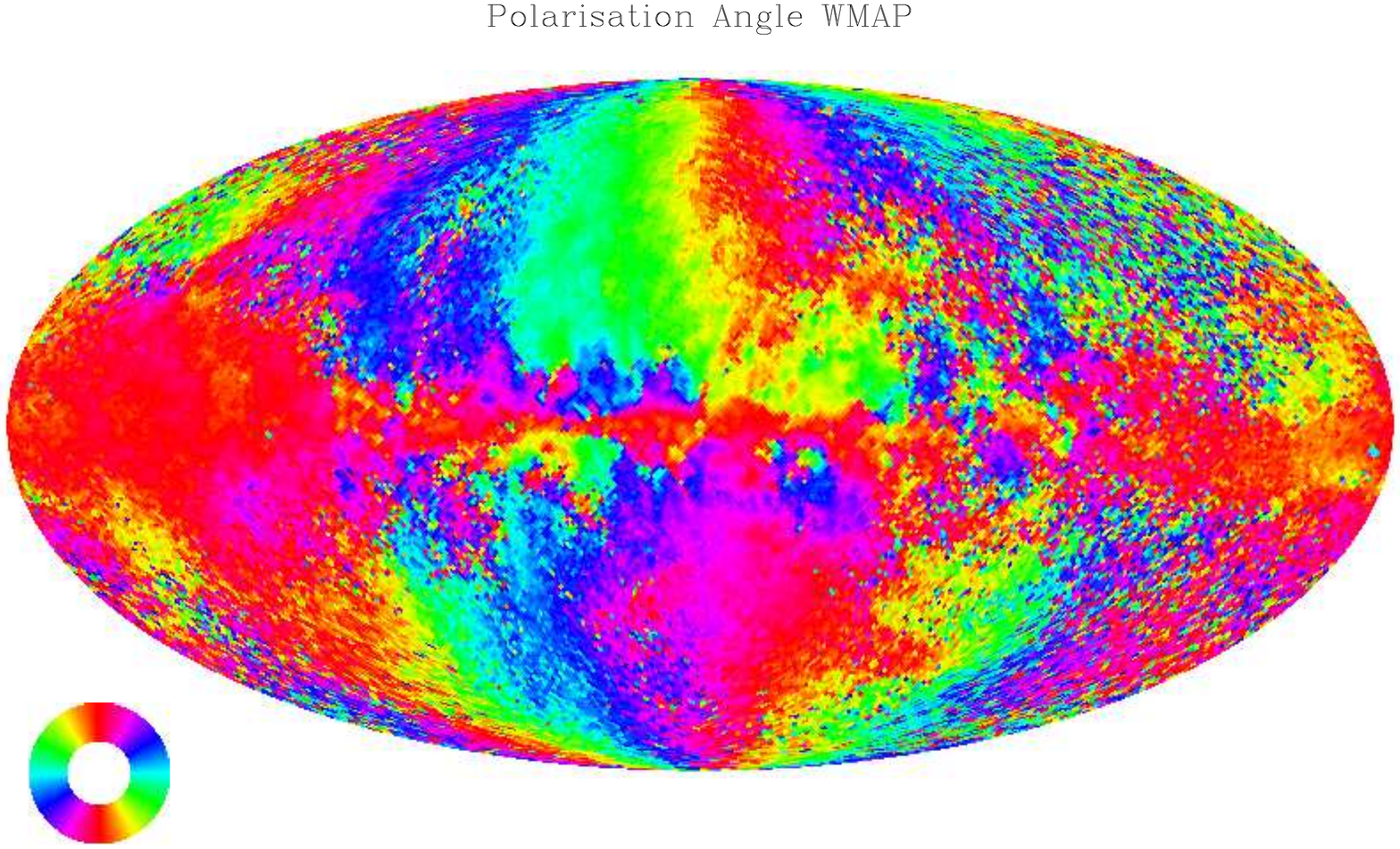}} \\
 \end{tabular}
 \caption{Polarisation amplitude ($P=\sqrt{Q^2+U^2}$) in antenna
   temperature ($\mu K$) ({\sl top}) and angle ({\sl bottom}) in
   Galactic co-ordinates for the synchrotron emission template (LSA
   GMF model) at 23 GHz ({\sl left}) and for \WMAP\ 23GHz ({\sl
     right}). The polarisation angle colour coding ranges from 0 to
   180 degrees.}
\label{fig:pol}
 \end{center}
\end{figure*}

\section{Polarised Synchrotron Emission}
\label{sec:synch}

Diffuse synchrotron emission is one of the dominant Galactic
foregrounds for CMB observations.  The synchrotron radiation arises
when electrons with large relativistic energies are accelerated in the
GMF.  The frequency dependence of synchrotron emission depends on the
energy spectrum of these cosmic-ray electrons, as well as the intensity
of the GMF (see e.g. ~\cite{1979rpa..book.....R}).  An ensemble of
relativistic electrons with a power law distribution in energy
produces a synchrotron emission spectrum that is another power law
(~\cite{1994hea2.book.....L}).  At GeV energies, where radio
synchrotron emission peaks, the index of the power law is expected to
have a range between $\beta \sim -3.5\rightarrow
-2.5$~\citep{1979rpa..book.....R} in inferred antenna temperature or,
equivalently, $\alpha \sim -1.5\rightarrow
-0.5$ in specific intensity.  Since the spectral index has been
seen to vary with position across the sky~\citep{2007ApJS..170..288H},
such a power law in antenna temperature describing synchrotron emission
is only an approximation.  In addition, the highest energy electrons
lose energy more quickly resulting in a gradual steepening in the
power law index at the higher
frequencies~\citep{2003ApJS..148...97B}.

At microwave frequencies polarised foreground emission is dominated by
polarised synchrotron and thermal dust that are both sensitive to the
coherent GMF~\citep{2007ApJS..170..335P}.  The dominant emission at
lower frequencies is polarised synchrotron radiation. \WMAP\
measurements from its lower frequency bands provide important
constraints on polarised synchrotron emission. Synchrotron emission is
linearly polarised with direction perpendicular to the projection of
the GMF on the plane of the sky (see for
example~\cite{1979rpa..book.....R}).  The degree of synchrotron
polarisation depends greatly on position on the sky and observing
frequency. Changes in magnetic field direction along the line-of-sight
leads to a depolarization effect, reducing the fractional polarisation
degree of synchrotron emission.  At frequencies lower than
$\sim$~1~GHz, depolarization is significant and hence synchrotron
polarisation is as low as few tens of
percent~\citep{1984A&A...135..238S}. At CMB frequencies,
depolarisation is minimal, with the degree of synchrotron polarisation
being as high as 30 to 50 \% in some galactic structure.

Free-free emission and spinning dust are also
thought to contribute to the foreground signal in total intensity over
a range of frequencies. For example, anomalous microwave emission
around 20GHz has been found in \WMAP\ data, with suggestions that this
is more likely due to spinning dust emission than a flat synchrotron
component \citep{2011arXiv1112.0432P}. However
spinning dust and free-free emission are not thought to be
significantly polarised and their impact on final estimates of e.g. the
tensor-to-scalar ratio $r$ is minimal \citep{2012arXiv1203.0152A} at
about the 1\% level.

Aside from the spatial dependence of the polarisation angle and
amplitude, significant uncertainties remain in the frequency modelling
of synchrotron intensity. For example, \citet{2003ApJS..148...97B}
argue that at higher Galactic latitudes (in the halo) the spectral
index $\beta \sim -3$ while in the Galactic plane (near star forming
regions) $\beta \sim -2.5$. This results in differences in the
observed structure between \WMAP\ $K$-band at 23GHz and the Haslam map at
408MHz \citep{HaslamII} as regions with flatter spectral
index become more important at higher frequencies. However
\citet{2008arXiv0802.3345M} find a lower range of variation of the
spectral index. This work focuses only on the polarisation fraction and
orientation due to the assumed GMF model. We assume a simple frequency
scaling of the Haslam template with a single spectral index multiplying
the internally modelled polarisation fraction to provide
morphologically realistic templates. A more detailed, possibly pixel
dependent, frequency rescaling can always be introduced by rescaling
the template obtained.

A number of other studies aimed at modelling foreground emission at
microwave frequencies have been carried out (see
e.g. \citet{2010arXiv1003.4450F}, \citet{2007ApJS..170..335P}). An
extensive study modelling different foregrounds in both intensity and
polarisation over a large range in angular scales is the Planck Sky
Model (PSM) \citep{2012arXiv1207.3675D}\footnote{The PSM description
  was released in the interim betwen \FGPol I and this work. The PSM
  template and model are available on a restricted basis and detailed
  comparisons will be described in a future publication.}. It includes
detailed modelling of Galactic diffuse emission, including synchrotron
and thermal dust emission as well as free-free, spinning dust and CO
lines. It also includes information on Galactic HII regions,
extragalactic radio sources and several other sources of emission.

At present the best template of polarised Galactic synchrotron
emission is that provided by the \WMAP\ $K$-band (23GHz) whilst the
intensity has been well measured (free from CMB contamination) at
408MHz by the Haslam all sky survey. As detailed below we use the
Haslam maps to introduce detailed morphology in our templates since
the \WMAP\ Maximum Entropy Method (MEM) maps still contain a
significant noise residual due to CMB contamination at the smoothing
scale adopted (1$^\circ$). We compare the templates obtained here with
the \WMAP\ synchrotron and dust maps obtained through their Monte
Carlo Markov Chain (MCMC) best--fit procedure
\citep{2011ApJS..192...15G}. Similar templates
\citep{2010arXiv1003.4450F} have been compared with \Archeops\ maps
over a limited fraction of the sky at 353 GHz
\citep{2004A&A...424..571B} but these maps are not publicly available.

\section{Model}\label{sec:synchmodel}

\subsection{Galactic Magnetic Field Model}\label{sec:GMF}

A  number  of  magnetic  field   models  were  compared  in  \FGPol  I
\footnote{We refer the reader to \FGPol I for a detailed discussion of
  GMFs on  large and small scales.}.  Here we limit the  choice to the
Logarithmic     Spiral    Arm     (LSA)     model    introduced     by
\cite{2007ApJS..170..335P} for use in  modelling the \WMAP\ data.  The
model is defined as
\begin{align}
  B_{\rho} &= -B_0 \sin{\left(\psi_0 + \psi_1
      \ln{\frac{\rho}{\rho_{\rm W}}}\right)}\cos{\chi}\,,  \nn\\
  B_{\Phi} &= -B_0 \cos{\left(\psi_0 + \psi_1
      \ln{\frac{\rho}{\rho_{\rm W}}}\right)}\cos{\chi}\,, \nn\\
  B_{z} &= B_0 \sin{\chi}\,,
\label{eq:lsamodel}
\end{align}
where $\rho$, $\Phi$ and $z$ are Galacto-centric cylindrical
co-ordinates with $\Phi$, the cylindrical longitude, measured from the
direction of the Sun, $\chi = \chi_0 \tanh(z/z_0)$ parametrizes the
amplitude of the $z$ component and $z_0=1$~kpc.  The field amplitude
is set to $B_0 = 3 \mu$G, and we take the distance between the Sun and
the Galactic center to be $8$~kpc.  Best-fit parameter values obtained
by fits to the \WMAP\ $K$-band field directions are $\psi_0=27$
degrees, $\psi_1=0.9$ degrees, and $\chi_0=25$ degrees. The radial
scale is set to $\rho_{\rm W}=8$~kpc and the scale height is set to
$z_0=1$~kpc.

Although we focus on large angular scales we also include a
small-scale random component in our GMF model by adding a realisation
of a Kolmogorov turbulence field with a one-dimensional Kolmogorov
energy spectral index of $-5/3$. An injection scale of $100$~pc is
chosen for the turbulent realisation with a negligibly small
dissipation scale compared to the resolution scale.

\subsection{Cosmic Ray Density Distribution}
 \label{sec:crdensity}

 To model synchrotron emission a three-dimensional model of the
 distribution of cosmic rays in the Galaxy is required. The
 large-scale spatial distribution of the cosmic rays is modelled
 through its density, $n_{\rm cr}$, and is thought to follow the same
 form as the dust distribution with modified radial and scale heights
 \citep{2007ApJS..170..335P},
\begin{equation}
n_{\rm cr} = n_0 \exp{\left(-\frac{\rho}{\rho_{{\rm cr}}}\right)}
\sech^2{\left(\frac{z}{z_{{\rm cr}}}\right)}\,.
\label{eq:chapthree7}
\end{equation}
where the height and radial scales are set to $z_{{\rm cr}} = 1$~kpc
and $\rho_{{\rm cr}} = 5$~kpc. These parameter values were chosen for
the \WMAP\ analysis of \citet{2007ApJS..170..335P} following work by
\citet{2001ApJ...556..181D}.

\subsection{Total Intensity}

Our method aims to predict the polarisation amplitude and angle based
on a chosen GMF model. The choice of total intensity for our templates
is therefore external to the model and can be set to reflect any
existing template. For synchrotron emission we chose to scale the
point source corrected Haslam all-sky survey using a single power law
in antenna temperature for simplicity. This scaled map is also
multiplied by the internally modelled polarisation fraction template
to produce Stokes parameter maps with realistic morphology.

The Haslam template in brightness temperature, with a resolution of
$0.85^\circ$, is scaled to microwave frequencies using a spectral
index $\beta_s =-3$. Although the map may contain residual
contamination by free-free emission (see e.g.
\citet{2003MNRAS.341..369D}) we assume it is dominated by the
synchrotron component. The templates can be rescaled using any choice
of templates in future.

\subsection{Stokes Parameters}
\label{sec:stokes}

The direction and degree of polarisation from synchrotron emission are
highly dependent on the Galactic magnetic field. To model these we
integrate along lines-of-sight using the GMF outlined in
section~\ref{sec:GMF}. The full-sky maps presented here were obtained
using a one-dimensional realisation of the small-scale turbulent
field. When producing smaller patches that require much fewer
lines-of-sight at a given resolution we model the small scale
turbulence as a full three-dimensional random realisation which
preserves the spatial correlations implied by the Kolmogorov spectrum.

The total GMF model is made up of a sum of large-scale (ls) and
small-scale (ss) components with
\begin{align} 
B_r=B_{r,\rm {ss}}+B_{r,{\rm ls}}\,,\nn \\
B_\theta=B_{\theta,{\rm ss}}+B_{\theta,{\rm ls}}\,, \nn \\
B_\phi=B_{\phi, {\rm ss}}+B_{\phi,{\rm ls}}\,, 
\end{align}
where $r$, $\theta$, and $\phi$ are now Solar-centric spherical polar
co-ordinates. The polarisation at each point along the line-of-sight
$\hat r$ is determined by the perpendicular field components,
$B_\theta$ and $B_\phi$.

The Stokes parameters for the synchrotron model are then projected out
from the three-dimensional model using the appropriate line-of-sight
integrals,
\begin{align}
  I_{\rm model}(\theta, \phi) &=\epsilon(\nu) \int_0^{r_{\max}} n_{{\rm cr}}(\vect{r})(B_{\phi}(\vect{r})^2+B_{\theta}(\vect{r})^2) \, \ud r\,, \nn\\
  Q_{\rm model}(\theta, \phi) &=\epsilon(\nu)\int_0^{r_{\max}} n_{{\rm
      cr}}(\vect{r}) p
  \frac{(B_\phi(\vect{r})^2 - B_\theta(\vect{r})^2)B_r^2}{{B^2}} \, \ud r \,, \nn\\
  U_{\rm model}(\theta, \phi) &= \epsilon(\nu)\int_0^{r_{\max}}
  n_{{\rm cr}}(\vect{r}) p
  \frac{2B_\phi(\vect{r})B_\theta(\vect{r})B_r^2}{{B^2}}\, \ud r\,,
\label{}
\end{align}
where $B^2=B_r^2+B_{\phi}^2+B_{\theta}^2$ and $\epsilon$ is the
emissivity as a function of frequency, $\nu$. As with the dust
templates, we conform to the default convention applied in the
\healpix\footnote{See http://healpix.jpl.nasa.gov} package
\citep{2005ApJ...622..759G} regarding the sign of $U$.

Having computed the line-of-sight integrals for the Stokes parameters
we calculate maps of the polarisation direction, $\gamma$, and degree,
$P$, given by
\begin{align}
  P(\theta, \phi) &= \frac{\sqrt{Q_{\rm model}^2 +
      U_{\rm model}^2}} { I_{\rm model}}\,, \nn\\
  \gamma(\theta, \phi) &= \frac{1}{2}
  \arctan{\left(\frac{U_{\rm model}}{Q_{\rm model}}\right)}\,.
\end{align}

The final synchrotron template at frequency $\nu$ is then obtained by
scaling with the Haslam template
\begin{align}
 I^\nu_{{\rm sync}} (\theta, \phi) &= I^\nu_{{\rm Has}}(\theta,\phi)\,,\nn\\
 Q^\nu_{{\rm sync}}(\theta, \phi) &= I^\nu_{{\rm Has}}(\theta,\phi)\,P(\theta,\phi) \,\cos\left(2\, \gamma(\theta,\phi)\right)\,, \nn\\
 U^\nu_{{\rm sync}}(\theta, \phi) &= I^\nu_{{\rm Has}}(\theta,\phi)\,P(\theta,\phi)\,\sin\left(2 \,\gamma(\theta,\phi)\right)\,,
\label{}
\end{align}
where $I_{{\rm Has}}^\nu$ is the total intensity of the Haslam map
extrapolated to frequency $\nu$.

\section{Maps}
\label{sec:maps}

Figure~\ref{fig:iqu} shows $Q$ and $U$ Stokes parameter maps at 23GHz for
the whole sky arising from the model with their amplitudes scaled such
that the polarisation amplitude corresponds to that of the \WMAP\
counterpart (also shown) when averaged over the maps.  The morphology
of the polarisation agrees well with the observations with the most
visible difference being on scales of a few degrees where the \WMAP\
estimates are dominated by residual noise. 

The resolution of the \healpix\ maps is $N_{\rm side}=1024$ but the
polarisation information is based on a line-of-sight integral at an
angular resolution of $N_{\rm side}^{\rm P}=128$, corresponding to
roughly $\ell \sim 500$ in multipole space. We integrate along
lines-of-sight to the centre of all \healpix\ pixels at a given
$N_{\rm side}^{\rm P}$ from zero out to a maximum distance $r_{\max}$
of 30,000~pc, with discretisation steps of $0.1$~pc. $N_{\rm
  side}^{\rm P}$ is less than or equal to $N_{\rm side}$ of the total
intensity Haslam map.

The LSA model is used for the Galactic magnetic field model with the
same parameters as in \FGPol I. The small scale field is modeled as
Kolmogorov turbulence in 1D for large patches of sky, with
a power spectrum of $\mathcal{P}(k) \propto k^{-(2+3\,N_d)/3}$ where
$N_d$ is the number of spatial dimensions of the realisation and $k$
is the magnitude of the wavevector. All full-sky templates presented
here make use of the 1D approximation along the line-of-sight for the
small scale turbulent component of the GMF. Small patch templates
discussed below are produced with full 3D realisations of the field in
the (smaller) volume probed by the reduced
coverage\footnote{Figure~\ref{fig:iqu} can be compared with Figure 2 in
\citet{2010arXiv1003.4450F}}. 

Figure~\ref{fig:pol} shows maps of $P$ and $\gamma$ obtained using
this choice of resolution and modelling of large and small scales. For
comparison, maps of $P$ and $\gamma$ for the \WMAP\ 23GHz MCMC
template are also plotted. Differences between the templates and
observations are mostly due to noise but there are also obvious
differences in the morphology along the galactic plane and around the
largest Galactic features such as the Galactic centre and North and
South Galactic Spurs. Some of these differences are related to our
choice of total intensity template which uses the Haslam maps at 408
MHz. A comparison between the scaled Haslam map and synchrotron
templates obtain via the differencing of \WMAP\ $K$ and $Ka$ bands
were discussed in \citet{2011ApJS..192...15G}. Below we quantitatively
compare the broad features of both synchrotron and dust full--sky
templates with the corresponding \WMAP\ MCMC best--fit maps.

\begin{figure}
 \begin{center}
\includegraphics[width=6.5cm,angle=270,trim=0cm 0cm 0cm 0cm,clip]{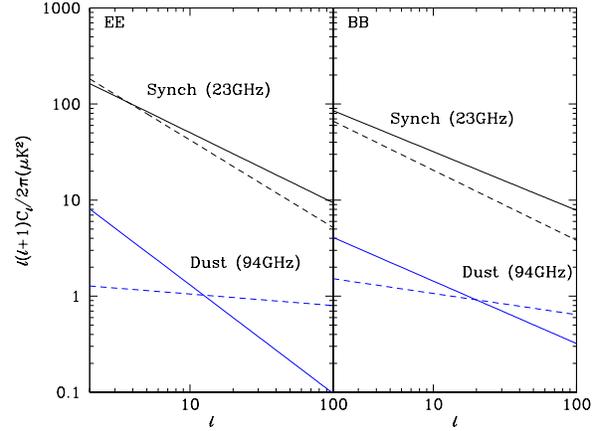}
\caption{$EE$ ({\sl left}) and $BB$ ({\sl right}) fits to the
  foreground angular power spectra for the \WMAP\ MCMC synchrotron
  23GHz and dust 94GHz maps (solid line) compared to the \FGPol\
  synchrotron model and dust model (dashed line). The polarisation
  amplitude of the model templates is matched to the \WMAP\ MCMC
  templates as described in the text. The fit values used are given in
  Table~\ref{table:fits}. These were calculated from maps masked with
  a union of the \WMAP\ P06 mask and the MCMC flagged pixels.}
\label{fig:spectral_comp}
\end{center}
\end{figure}

\begin{table}
\centering
\begin{tabular}{|c c c|}
  \hline
  Component & A [$\mu K^2$] & m \\[0.5ex]
  \hline
  \WMAP\ MCMC & & \\
Synchrotron $EE$ & 306 $\pm$ 95& $-0.91$ $\pm$ 0.11\\
Synchrotron $BB$ & 144 $\pm$ 44& $-0.87$ $\pm$ 0.12\\
Dust $EE$ & 12.9 $\pm$ 6.4& $-1.06$ $\pm$ 0.24\\
Dust $BB$ & 6.12 $\pm$ 3.7& $-0.83$ $\pm$ 0.29\\
\hline
  \FGPol\ & & \\
Synchrotron $EE$ & 343 $\pm$ 91 & $-1.05$ $\pm$ 0.09\\
Synchrotron $BB$ & 110 $\pm$ 29 & $-0.73$ $\pm$ 0.08\\
Dust $EE$ & 1.38 $\pm$ 0.33 & $-0.12$ $\pm$ 0.07 \\
Dust $BB$ & 1.70 $\pm$ 0.37 & $-0.22$ $\pm$ 0.06 \\
\end{tabular}
\caption{Foreground power law + white noise fits of \WMAP\ MCMC and
  \FGPol\ template spectra outside the combination of P06 mask and MCMC flagged pixels. There is good agreement in both $EE$ and
  $BB$ for synchrotron
  between the \FGPol\ template and the \WMAP\ MCMC synchrotron
  component map. The \FGPol\ dust template shows a significantly
  shallower spectrum than the \WMAP\ MCMC component map indicating
  relatively more structure at large angular scales. }
\label{table:fits}
\end{table}

\section{Comparison with \WMAP\ templates}
\label{sec:wmap_comp}
The \WMAP\ satellite observations provide full--sky maps of temperature and
polarisation in five frequency bands between 23GHz and 94GHz
\footnote{http://lambda.gsfc.nasa.gov} \citep{Jarosik:2010iu}. The
polarisation maps contain important information on Galactic foreground
emission and hence provide an important test of our model of Galactic
synchrotron radiation. Synchrotron radiation dominates the measured
signal in the lower frequency bands, and we use the best fit \WMAP\
23GHz synchrotron templates generated by an MCMC fit for comparison
with our model. Thermal dust emission dominates higher frequency
bands, and for comparison with our dust model we use the 94GHz dust
templates generated from the MCMC best fit values.

We use the \WMAP\ MCMC `base' fit which includes three power law
foregrounds: dust, synchrotron and free-free emission as well as a
contribution from CMB. These maps are smoothed at a scale of $1^\circ$
and have been downgraded to $N_{\rm side}=64$ before the MCMC fit is
carried out.
The \WMAP\ team performed a combined MCMC fit to their five bands at a
resolution of $N_{\rm side}=64$. The pixel noise is calculated at
$N_{\rm side}=512$ and downgraded to $N_{\rm side}=64$. 

\FGPol\ templates are normalised to the polarisation amplitude
$P=\sqrt{Q^2+U^2}$ of the full sky \WMAP\ MCMC template with no
masking or smoothing. The model maps are generated at $N_{\rm
  side}=1024$ for the large scale resolution and $N_{\rm side}^{\rm
  P}=128$ for the small scale line-of-sight resolution, then smoothed
with a Gaussian beam of $1^{\circ}$ and degraded to $N_{\rm side}=64$.

Angular power spectra $C_\ell^{XX}$ for $XX\equiv TT, EE$, and $BB$ of
both templates were calculated after masking with the P06 polarisation
mask \citep{2007ApJS..170..335P} combined with a mask of pixels
flagged by the \WMAP\ MCMC process. The spectra are corrected for sky
fraction $f_{\rm sky}$ effects and for their respective pixel and beam
smoothing functions and then fit for a power law in multipole $\ell$
with an additional white noise component
\begin{align}
\label{eq:cl}
 \frac{\ell(\ell+1)}{2\pi}C_\ell &= A\ell^m+\ell(\ell+1)N^2\,
\end{align}
where $A$ is the amplitude of the foreground component, $m$ is the
index and $N$ is the noise amplitude. This procedure is similar to
that carried out by \citet{2010arXiv1001.4555G}. Although the scatter
at large angular scales is non-Gaussian and somewhat correlated by the
sky cut we adopt a very simple assumption for sample variance in the
power spectra by disregarding correlations between multipoles and
assuming a Gaussian scatter given by the sample variance for each
$C_\ell^{XX}$. The \WMAP\ derived fits also make use of diagonal
Fisher error values.

The fits allow us to quantify the scaling of the angular power
spectrum for both templates as a function of multipole $\ell$ whilst
allowing for any residual noise and/or pixelisation effects. They can
also be used as a quick guide for the level of foreground
contamination at different frequencies on large angular scales either
on the full sky or on small patches.  Figure~\ref{fig:spectral_comp}
shows the resulting power law fits in both $C_\ell^{EE}$ and
$C_\ell^{BB}$ for the P06 masked \FGPol\ and \WMAP\ MCMC synchrotron
templates at a frequency of 23 GHz. Also included are the results of
the same procedure applied to the dust \FGPol\ and \WMAP\ MCMC
templates at 94 GHz.

The fit values, excluding the noise amplitudes, can be found in
Table~\ref{table:fits}. The synchrotron templates agree well with the
\WMAP\ MCMC maps in both amplitude and angular dependence whereas
there are significant differences between the \FGPol\ dust template
and the \WMAP\ MCMC map at 94 GHz.

\begin{figure}
 \begin{center}
   \includegraphics[ width=6cm,height=7.5cm,angle=270,trim=0cm 0cm 0cm 0cm,clip]{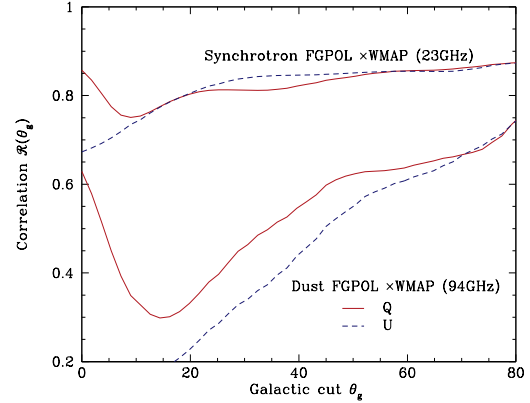}
   \caption{Comparison of the correlation coefficient $\cal R$ between
     the \FGPol\ maps and \WMAP\ best--fit templates for both
     synchrotron and dust. In order to look at the correlation on large scales we smooth these maps to a resolution of $10^{\circ}$. The correlation is calculated by including
     areas outside a range of cuts in Galactic latitude. The trend
     shows that both dust and synchrotron models are highly correlated
     with the best-fit maps at high Galactic latitudes whilst the
     correlation falls rapidly at low galactic latitudes for the dust
     comparison. The \WMAP\ best--fit dust map however has an even
     larger noise residual than the synchrotron map and this is
     expected to reduce the correlation significantly.}
\label{fig:corr}
 \end{center}
 \end{figure}

We also attempt to quantify the level of correlation between the
\WMAP\ MCMC maps and \FGPol\ templates. We do this using two separate
measures. The first analyses the level of pixel-to-pixel correlation
between maps calculated for the area of the sky outside a given
galactic latitude cut. The correlation coefficient ${\cal R}$ is
given by
\begin{align}
  {\cal R}(\theta_g) &= \frac{\sum_p({\cal W}_p - \widehat {\cal
      W}_p)({\cal F}_p - \widehat {\cal F}_p)}{\sqrt{\sum_p({\cal W}_p
      - \widehat {\cal W}_p)^2\sum_p({\cal F}_p - \widehat {\cal F}_p)^2}}\,.
\end{align}
where ${\cal W}$ and ${\cal F}$ are the $I$, $Q$, or $U$ Stokes values
of the \WMAP\ and \FGPol\ maps respectively and the index $p$ sums
over all pixels outside the cut at latitude $\pm \theta_g$. The
result, for both dust and synchrotron $Q$ and $U$ Stokes parameters is
shown in Figure~\ref{fig:corr}. The analysis shows that both dust and
synchrotron templates are highly correlated with the \WMAP\ best-fit
foreground templates at high Galactic latitudes. Whilst this is also
true for synchrotron at low Galactic latitudes, the dust model fails
to reproduce the observed morphology well at latitudes below $\sim$ 30
degrees. This is not surprising since thermal emission by dust
particles is more susceptible to the detailed structure in the
Galactic disk with even large angular scales being influenced by
turbulence and/or existence of individual clouds. 

\begin{figure}
  \begin{center}
    \includegraphics[height=6cm, width=8cm,angle=0,trim=1cm 12cm 1cm 3cm,clip]{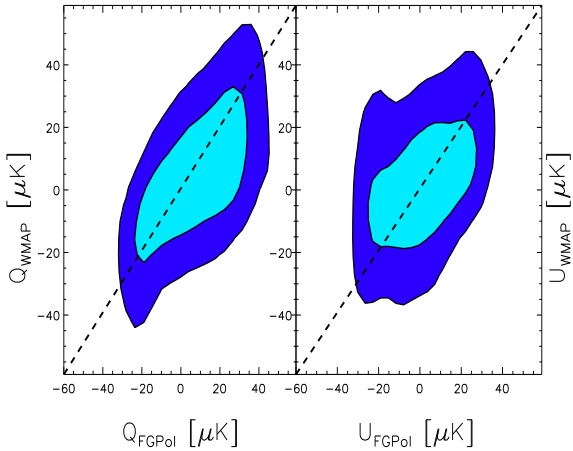}
    \caption{Comparison of the scatter between the \FGPol\ maps and
      \WMAP\ best--fit templates for synchrotron. This is done for the
      full--sky templates including pixels outside the combination of
      the P06 polarisation mask and \WMAP\ MCMC flagged pixels. The
      dust correlation density is omitted as it is dominated by the
      variance in the \WMAP\ map due to residual noise.}
    \label{fig:scatter}
  \end{center}
\end{figure}

We also look at the correlation in terms of scatter of the pixel
values in $Q$ and $U$ Stokes parameters for the \WMAP\ MCMC maps
versus the \FGPol\ templates in both dust and synchrotron. All pixels
outside the P06 mask and MCMC flagged pixels are included and the
scatter density is shown in Figure~\ref{fig:scatter} as two contours
encompassing 68\% and 95\% of pixels. We only show the synchrotron
correlation density since the dust one is found to be dominated by the
larger \WMAP\ variance due to residual noise and is not informative.

\section{Foreground amplitudes in sub--orbital sky patches}
\label{sec:suborb}

We also examine the amplitude of foreground contamination in smaller
sky areas being targeted by a sample of three currently operating or
planned sub-orbital experiments; \ebex\ \citep{2010SPIE.7741E..37R},
\spider\ \citep{2010SPIE.7741E..46F} and the \biceptwo\ and \keck\
\citep{2010arXiv1009.3685O} arrays which observe the same
field. Angular resolution and sensitivity for the three experiments
are varied but they are all targeting the detection of $BB$ power
either on large angular scales that are free of lensing effects or, as
in the case of \ebex, on smaller angular scales where the lensing
effect dominates the $BB$ signal. 

\begin{figure}
  \begin{center}
    \includegraphics[width=9cm,angle=0,trim=0cm 9cm 0cm
    0cm,clip]{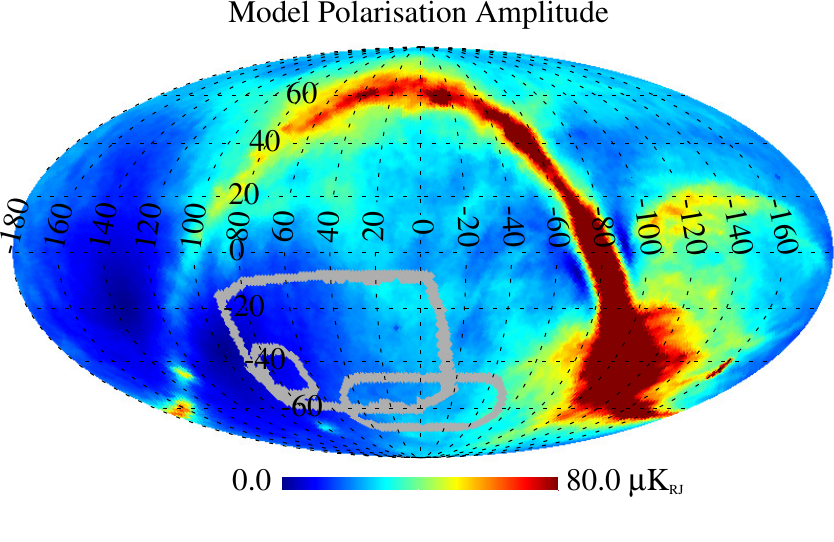}
    \caption{Comparison of patches targeted by a variety of
      sub-orbital experiments. \spider\ targets the largest area
      ($f_{\rm sky}\sim 0.1$), \ebex\ targets the smallest patch
      ($f_{\rm sky}\sim 0.01$) and \biceptwo/\keck\ targets the
      southern most patch ($f_{\rm sky}\sim 0.03$).}
    \label{fig:combinedmasks}
  \end{center}
\end{figure}

We generate high resolution
templates with full three-dimensional modelling of the turbulent
small-scale GMF over the regions of expected coverage for the three
experiments. For the dust templates, the $Q$ and $U$ components are normalised so that the average polarisation fraction outside the area defined by the WMAP P06 mask is 3.6\%. The coverage areas are outlined in Figure~\ref{fig:combinedmasks} with
\spider\ targeting the largest area with $f_{\rm sky}\sim 0.1$, \ebex\
targeting the smallest patch contained in the \spider\ area with
$f_{\rm sky}\sim 0.01$ and \biceptwo/\keck\ targeting the southern
most patch with $f_{\rm sky}\sim 0.03$.

In order to compare the relative contamination by foregrounds in
relation to the relevant signal we analyse the angular power spectrum
$C_\ell^{BB}$ for both dust and synchrotron. Due to the uncertainties
involved in predicting the small scale signal we focus on large
scales only with templates smoothed to a common resolution of
$1^\circ$ and rely on extrapolating a power law to scales larger than
$\ell \sim 200$ in comparing with the expected signal.

The analysis on small areas of the sky such as these is complicated by
the significant correlation induced by the cut on spherical harmonic
coefficients. The high level of correlation would result in
significant biases if the same power law fitting procedure as used in
Section~\ref{sec:wmap_comp} were carried out. To avoid this problem we
estimate the overall amplitude of foreground contamination by
averaging in pixel space assuming a fixed power law in $\ell$
corresponding to our previous near full-sky analysis. 

In practice we calculate the variance in both $Q$ and
$U$ for each patch and assume a relation between the variance and
angular power spectrum of the form
\begin{equation}
\sigma^2 = \frac{1}{4\pi}\sum_{\ell=2}^{\ell_{\rm max}} (2\ell+1)C_{\ell}B^2_{\ell}(\theta_s)\,,
\label{eq:sigma0}
\end{equation}
with the signal angular power spectrum modelled as $C_\ell=A\,
\ell^{m}$ in accordance with (\ref{eq:cl}) and with index $m$ set to
the corresponding near full-sky best-fit value (see
Table~\ref{table:fits}). We take $\ell_{\rm max}=128$ and model the
smoothing $B_\ell$ applied to the templates as a Gaussian beam with
FWHM $1^\circ$ multiplied by the pixel window function at the working
{\tt HealPix} resolution $N_{\rm side}=64$. We then invert the
relation (\ref{eq:sigma0}) to obtain an `average' polarisation angular
power spectrum amplitude $A$, effectively assuming that power is
equally distributed between $EE$ and $BB$.

\begin{figure}
  \includegraphics[width=6.5cm,angle=270,trim=0cm 0cm 0cm 0cm,clip]{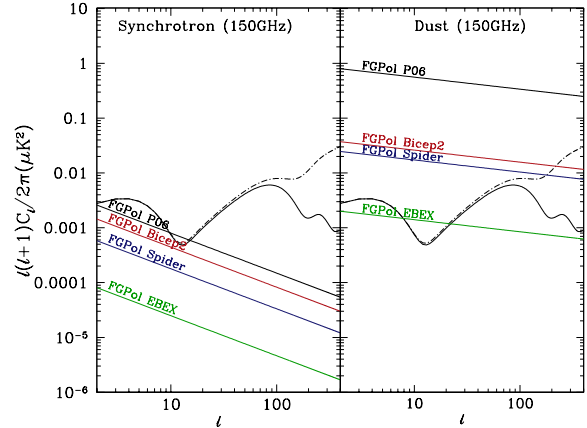}
  \caption{Foreground amplitude for synchrotron (left panel) and dust
    (right panel) calculated from patches targeted by various
    suborbital experiments compared with theoretical $EE$ and $BB$
    spectra for $r=0.1$. The index of the power law used is from the
    corresponding near full sky power spectrum fits. Also shown are
    the near full sky best fit spectra for the \FGPol\ synchrotron
 template from Table~\ref{table:fits} along with best fit spectra to the dust templates with $Q$ and $U$ normalised so that the polarisation fraction is 3.6\% outside the WMAP P06 mask. The amplitudes
    were calculated from our \FGPol\ dust and synchrotron templates at
    150GHz calculated from maps generated at $N_{\rm side}=1024$ for
    the large scale resolution and $N_{\rm side}^{\rm P}=128$ for the
    small scale line-of-sight resolution, which are then smoothed to
    $1^{\circ}$ and downgraded to $N_{\rm side}=64$. Units are $\mu
    K_{\rm CMB}$.}
  \label{fig:patchforegrounds}
\end{figure}

The results are summarised in Figure~\ref{fig:patchforegrounds} for a
single reference frequency of 150 GHz as this is being included as an
observing frequency in all experiments being considered. The model
power spectra for each patch are shown in thermodynamic temperature in
order to compare directly with the expected $BB$ signal for a
tensor-to-scalar ratio $r=0.1$. Both primordial and lensing
contribution to the $BB$ signal are shown. 

The amplitude of foreground contamination varies by roughly an order
of magnitude between the area targeted by different experiments. In
particular the area targeted by \ebex\ seems to be very clean with the
foreground signal reduced by an order of magnitude compared to the
areas targeted by \spider\ and \biceptwo/\keck. This agrees visually with the
impression given in Figure~\ref{fig:combinedmasks}.

\section{Conclusions}
\label{sec:conclude}
We have presented templates for polarised emission from synchrotron
radiation within our Galaxy using a 3D model of the Galactic magnetic
field and cosmic ray density distribution. From this model, maps of
polarisation amplitude and angle are calculated which are then
combined with total intensity measurements from the Haslam 408MHz
all-sky radio continuum survey to provide template maps.

We have compared the \FGPol\ templates obtained from this model with
data from the \WMAP\ satellite for both synchrotron and dust
emission. We find that the synchrotron template agrees qualitatively
with the observations whereas comparison of the dust template is
complicated by the large residuals present in the \WMAP\ estimates.

We have also looked at foreground contamination levels in patches that
will be targeted by upcoming experiments and found that our model
predicts significant differences of up to an order of magnitude in the
foreground contamination of different patches. The level of
contamination will dominate the ability of various experiments to
achieve their target sensitivity with respect to the $B$-mode signal
being searched for. 

As more polarisation data becomes available the comparison between the
model and observations will become more quantitatively precise. In
particular future Planck data releases will provide high
signal-to-noise $Q$ and $U$ maps at a number of frequencies and we
will be able to refine our model based on them. Indeed, in future, it
should be possible to learn much about the Galactic magnetic field
itself by fitting the (many) model parameters to actual data. This
will shed light on many aspects of our Galaxy's physical model that
are still poorly understood.

\section*{Acknowledgments}

We acknowledge Daniel O'Dea who provided the original GMF models this
work is based on. We also acknowledge Sasha Rahlin for an updated
\spider\ coverage mask and Cynthia Chiang for providing a \biceptwo\
coverage mask which we used to approximate the \keck\ coverage.  We
also thank \ebex\ team members Andrew Jaffe, Donnacha Kirk and Ben
Gold for providing us with a suitable \ebex\ mask. We acknowledge Clement Pryke for pointing out an error in amplitude in Figure 7 of a previous version of this paper. Caroline Clark is
supported by an STFC studentship. Carolyn MacTavish is supported by a
Kavli Institute Fellowship at the University of Cambridge.
Calculations were carried out on a facility provided by the Imperial
College High Performance Computing
Service\footnote{\url{http://www.imperial.ac.uk/ict/services/teachingandresearchservices/highperformancecomputing}}.

\bibliography{ForegroundsBib}
\label{lastpage}
\end{document}